\documentstyle[prb,aps,amsfonts,amssymb,epsf]{revtex}

\newcommand{\ucirc}{$\overcirc{\text{u}}$}
\newcommand{\tv}{t\hspace{-0.25em}'}
\begin{document}
\draft

\twocolumn[\hsize\textwidth\columnwidth\hsize\csname @twocolumnfalse\endcsname

\title{ Direct ab initio calculation of the dynamical matrix }

\author{
Marek Hy{\tv}ha and Anton\'{\i}n \v{S}im{\ucirc}nek
}
\address{
Institute of Physics of the Czech Academy of Science,
Cukrovarnick\'{a}~10, 162~53~Praha~6, Czech Republic
}

\date{\today}
\maketitle


\begin{abstract}
In this paper there is presented method for {\em ab initio} calculation
of the phonon spectra. The method is based upon a direct calculation of
the dynamical matrix via second derivatives of the total energy.
The pseudopotential technique in plane-wave basis set was used
to calculate the total energy within the local density approximation
(LDA).
For the change of the electron density there is derived the
self-consistent equation which is solved analytically with no use of
iterations.
In this paper the attention is paid only to non-metallic systems.
\end{abstract}


\vskip2pc]
\narrowtext

\section{Introduction}

Increasing accuracy of methods for calculation of the electronic structure of
solids enable us to obtain the variety of properties of the solid state.
In this paper, there is described new scheme for {\it ab initio} calculations
of phonon spectra in non-metallic systems.

The problem of phonons as the perturbation of the perfect crystal lattice
is used to be studied in different ways. The first one is the so-called direct
approach \cite{Ihm_et_al,Kunc} based upon the comparison of the properties
of unperturbed and perturbed states. The perturbed state is solved with use
of the supercell geometry. This method enable us to study both the linear
and nonlinear properties of the crystal lattice. Disadvantage of this
method is that for each phonon wavevector $\bf q$ we need a corresponding
supercell. Only a few phonons can be described by the supercell containing
only a few cells, and this is the reason why the use of this method
is limited only to few special long-wave phonons. Thus, this supercell
technique seems to be insufficient as there are a lot of quantities,
which must be computed by integrating over phonon frequencies of the whole
Brillouin zone.

The second way is to calculate the dynamical matrix which
can describe dynamical properties of solids with use of the linear
response methods.

In this paper there is presented method for {\it ab initio} calculation of
phonon spectra of non-metallic systems. At first the self-consistent electron
charge density is calculated. Then, the self-consistent variation of charge
density for an arbitrary phonon wave-length is calculated. After that,
that one is used for calculating of the dynamical matrix. The presented
scheme is restricted to non-metallic systems because the variation of the
numbers of states with respect to  perturbation in non-metallic systems
is almost zero. The situation in metals is more complex.

The presented method differs from the other linear
response methods, the dielectric matrix method \cite{Resta},
the method of Baroni {\it et al.} \cite{Baroni_et_al}, 
the method of King-Smith
and Needs \cite{King-Smith_Needs} and the method of 
Rignanese, Michenaud and Gonze 
\cite{Rignanese,Gonze}.
The presented method is based on direct calculation of the dynamical
matrix via the second derivatives of the total potential energy of the
crystal lattice.

Throughout the paper, the hartree units are used.

\section{Dynamical matrix}
In the harmonic approximation the theory 
of lattice dynamics can be formulated 
in terms of the dynamical matrix \cite{Maradudin}.
Then it can be written
\begin{equation} \label{dyn}
{\left[ {\it D}({\bf q})-\omega^2({\bf q}) \right]} {\it W} (\bf q)=0,
\end{equation}
where ${\it D}$ stands for the dynamical
matrix, ${\bf q}$ is the wavevector
of the phonon, $\omega$ is the frequency 
of the phonon and ${\it W}$ is the eigenvector
of the dynamical matrix which represents
the polarization of the phonon.
The dimension of the dynamical matrix is
${\it 3N\times3N}$, where ${\it N}$
stands for the number of atoms per cell.
Elements of the dynamical matrix 
can be expressed as \cite{Maradudin}
\begin{equation} \label{dynD}
D_{\alpha\beta}({\bf q},\mu\nu) =
\frac{1}{\sqrt{M_\mu M_\nu}}
\sum_{\bf h} A_{\alpha\beta}({{\bf h},\mu\nu})
\exp{(-i\bf{q\cdot h})} ,
\end{equation} 
where $M_\mu , M_\nu$ are 
masses of atom $\mu$ and $\nu$,
$\alpha$ and $\beta$ are Cartesian directions 
and $\bf{h}$ is vector of the
translation symmetry. Elements of the matrix
$\bf{A}$ are defined as
\begin{equation} \label{dynA}
A_{\alpha\beta}({\bf h},{\mu\nu})
\stackrel{def}{=}
\left.
\frac{\partial^2 V_L}
{
\partial {u}_{\alpha}^{\mu,0} 
\partial {u}_{\beta}^{\nu,{\bf h}}
}  
\right|_{{\bf u}=0} ,
\end{equation} 
where $V_L$ stands for the total potential energy of the crystal lattice,
$u_{\alpha}^{\mu,0}$
is a vector giving displacement of atom $\mu$ 
away from equilibrium in direction
of $\alpha$ axes. The $u_{\beta}^{\nu,{\bf h}}$ is a vector
of displacement away from equilibrium 
in the direction of $\beta$ axes
of atom $\nu$ in elementary cell 
with the translation 
vector $\bf{h}$ from the cell containing atom $\mu$. 
Defining the operator
\begin{equation} \label{oper}
\partial_{\mu,\alpha}^{q}
=
\left.
\frac{1}{\sqrt{M_\mu}}
\sum_{\bf h}
e^{-i{\bf q}\cdot{\bf h}}
\frac{\partial}{\partial u_{\alpha}^{\mu,h}}
\right|_{{\bf u}=0} ,
\end{equation}
one can write the
dynamical matrix as
\begin{equation} \label{part}
D_{\alpha\beta}({{\bf q},\mu\nu}) =
\partial_{\mu,\alpha}^{q*}
\partial_{\nu,\beta}^{q} V_L ,
\end{equation} 
where asterisk
 means the complex conjugated operator.

To solve the Eq.~(\ref{dyn}) we have 
to know elements of the dynamical matrix.
They can be calculated on the basis of
elements of matrix ${\it A}$. To calculate
elements of the matrix ${\it A}$ we
need the second derivatives of
energy $V_L$ for which
it can be written 
\begin{equation} \label{etot2}
V_L = V_{Ew} + E_{el} ,
\end{equation}
where $V_{Ew}$ is the Ewald ion-ion interaction energy
and ${E}_{el}$ stands for energy of the set of electrons.

To find the contribution of the $V_{Ew}$ to the dynamical matrix
is easy because the analytical expressions for that one are
well known \cite{Pickett} and therefore
in the rest of this paper we study only term $E_{el}$.

\section{The electron system energy}

Using the conventional density 
functional formalism \cite{Kohn-Sham}
in the pseudopotential framework,
the total energy
of the set of electrons is given by \cite{Ihm_et_al}
\begin{eqnarray} \label{E_el}
E_{el} &=&
E_{\varepsilon} 
- \frac{1}{2}
\int \rho({\bf r}) V_{c}({\bf r}) d^3 r
\nonumber \\ &-&
\int \mu_{xc}({\bf r})\rho({\bf r}) d^3 r + {E_{xc}[\rho]} ,
\end{eqnarray}
where $\rho$ is the charge density,
$V_{c}$ is the Coulombic potential, $\mu_{xc}$ is the
exchange and correlation potential and $E_{xc}$ stands for the
exchange-correlation energy. The $E_\varepsilon$ is the sum
of eigenvalues 
\begin{equation} \label{E_eps}
E_{\varepsilon} =
\sum_i
n_i
{\varepsilon}_i , 
\end{equation}
where ${\varepsilon}_i$ are eigenvalues
of the one-particle Kohn-Sham equation {\cite{Kohn-Sham}}
\begin{equation} \label{H_DFT}
(\stackrel{\wedge}{T} +
U + V_{c} + \mu_{xc})
\psi_{i}
= \varepsilon_{i} \psi_{i} .
\end{equation}
$\psi_{i}$ are self-consistent eigenfunctions,
$\stackrel{\wedge}{T}$ is operator
of kinetic energy and
$U$ stands for pseudopotentials
of the ionic sites. 
The exchange-correlation potential
$\mu_{xc}$ is given in the local density approximation by
\begin{eqnarray} \label{mu}
\mu_{xc}({\bf r}) &=&
\frac{\delta E_{xc}{[\rho({\bf r})]}}{\delta \rho({\bf r})} =
\frac{\delta \int{\epsilon_{xc}({\bf r})~ \rho({\bf r}) d^3 r}}
{\delta \rho({\bf r})}
\nonumber \\ &=&
\epsilon_{xc}({\bf r}) +
\frac{\delta \epsilon_{xc}({\bf r})}
{\delta \rho({\bf r})}
\rho({\bf r}).
\end{eqnarray}
The variation of $E_{el}$
can be then expressed  as
\begin{eqnarray} \label{E_elt}
\delta E_{el} = &&
\delta \sum_{i} n_{i} \varepsilon_{i} 
- \frac{1}{2}
\delta \int \rho({\bf r}) V_{c}({\bf r}) d^3 r
\nonumber \\ &-&
\delta \int
\frac{\delta \epsilon_{xc}({\bf r})}
{\delta \rho({\bf r})}
\rho({\bf r})
\rho({\bf r})
d^3 r .
\end{eqnarray}
The variation of the first term on the
right-hand side can be
written as
\begin{equation}
\delta \sum_{i} n_{i} \varepsilon_{i} 
=
\sum_{i} 
{[
n_{i} \delta \varepsilon_{i} 
+
\delta n_{i} 
(\varepsilon_{i}+ \delta \varepsilon_{i}) 
]}.
\end{equation}
In metallic systems the occupancy 
can change with an additional
potential while in the non-metallic 
systems the occupation number 
remains unchanged for small perturbation.
Thus, for the
non-metallic systems one can write
\begin{equation} \label{Eel_oneG}
\delta \sum_{i} n_{i} \varepsilon_{i}
=
\sum_{i} n_{i} \delta \varepsilon_{i} .
\end{equation}
Having used the perturbation series  
one obtains
\begin{eqnarray} \label{Mvar}
&& \delta E_{\varepsilon} =
\sum n_{v} \delta \varepsilon_{v} 
=
\sum_{v}
\langle{v}\mid 
\delta U +\delta V_{c} +\delta\mu_{xc}
\mid{v}\rangle 
\nonumber \\ && +
\sum_{v,c} \frac
{\langle{v}\mid 
\delta U +\delta V_{c} +\delta\mu_{xc}
\mid{c}\rangle 
\langle c \mid
\delta U +\delta V_{c} +\delta\mu_{xc}
\mid{v}\rangle} 
{{{\varepsilon}_{v}}-{\varepsilon}_{c}} 
\nonumber \\ && +
\Theta (\delta U +\delta V_{c} +\delta\mu_{xc}),
\end{eqnarray}
where $v$ and $c$ denote the occupied and 
unoccupied bands, respectively. 
The $\Theta$ stands for higher terms.
The first term can be expressed as
\begin{eqnarray} \label{first}
&&\sum_{v}
\langle{v}\mid 
\delta U +\delta V_{c} +\delta\mu_{xc}
\mid{v}\rangle
=
\sum_{v} 
\langle{v}\mid 
\delta U
\mid{v}\rangle
\nonumber \\ && +
\int \rho^{0}({\bf r}) \delta V_{c}({\bf r}) d^3 r
+
\int \rho^{0}({\bf r}) \delta \mu_{xc}({\bf r}) d^3 r .
\end{eqnarray}
Having used the expression for the variation of the Coulombic potential
\begin{equation} \label{varVc}
V_c({\bf r}) = \int \frac{\delta \rho({\bf r'})}{|{\bf r - r'}|} d^3 r',
\end{equation}
the difference of the second term of Eq.(\ref{E_elt}) and the second term
of Eq.(\ref{first}) can be written as
\begin{equation}
\rho^0({\bf r}) \delta V_{c}({\bf r})
-\frac{1}{2} \delta \left(
\rho({\bf r}) V_{c}({\bf r}) \right )
=
-\frac{1}{2} \int
\frac{ \delta \rho({\bf r}) \delta \rho({\bf r'}) }
{|{\bf r - r'}|} d^3 r'.
\end{equation}
Having expressed the exchange and correlation potential from Eq.(\ref{mu}),
the difference of the third term of Eq.(\ref{E_elt}) and the third term of
Eq.(\ref{first}) can be written as
\begin{equation} \label{var}
\rho^0 \delta\mu_{xc} -
\delta \left(\frac{\delta \epsilon_{xc}}{\delta \rho } \rho \rho \right) =
\delta \left( \rho^0 \epsilon_{xc} -
\left(\rho - \rho^0\right) \rho \frac{\delta \epsilon_{xc}}{\delta \rho}  \right).
\end{equation}
For the first order variation we can write
\begin{equation}
\frac{\delta}{\delta \rho}
\left( \rho^0 \epsilon_{xc} -
\left(\rho - \rho^0\right) \rho
\frac{\delta \epsilon_{xc}}{\delta \rho}
\right)_0
=0,
\end{equation}
and the second order variation can be written as
\begin{equation}
\frac{\delta^2}{\delta \rho^2} \left( \rho^0 \epsilon_{xc} -
\left(\rho - \rho^0\right) \rho
\frac{\delta \epsilon_{xc}}{\delta \rho}  \right)_0
=-\frac{\delta\epsilon_{xc}}{\delta\rho},
\end{equation}
where the subscript $0$ means that variations are taken for
$\rho = \rho^0$.
The variation of $E_{el}$
can be finally rewritten into the form
\begin{eqnarray} \label{E_elvar}
&& \delta E_{el} =
\sum_{v}
\langle{v}\mid 
\delta U
\mid{v}\rangle
\nonumber \\ &+&
\sum_{v,c} \frac
{\langle{v}\mid 
\delta U +\delta V_{c} +\delta\mu_{xc}
\mid{c}\rangle 
\langle{c}\mid 
\delta U +\delta V_{c} +\delta\mu_{xc}
\mid{v}\rangle}
{{{\varepsilon}_{v}}-{\varepsilon}_{c}} 
\nonumber \\ &-&
\frac{1}{2} 
\int \frac{\delta \rho({\bf r})
\delta \rho({\bf r}')}{|r-r'|} d^3 r d^3 r'
\nonumber \\ &-&
\int
\left(\frac{\delta \epsilon_{xc}({\bf r})}{\delta\rho({\bf r})}
\right)_0
\delta\rho({\bf r}) \delta \rho({\bf r}) d^3 r
\nonumber \\ &+&
\Theta (\delta U +\delta V_{c} +\delta\mu_{xc}).
\end{eqnarray}
For the first order variation of the exchange-correlation 
potential $\mu_{xc}$ we can write
\begin{equation} \label{var_mu}
\delta \mu_{xc} ({\bf r}) = 
\left ( 
\frac{\delta \mu_{xc} ({\bf r}) }{\delta \rho({\bf r})}
\right )_0
\delta \rho ({\bf r}).
\end{equation}
As it was already mentioned one needs to
calculate the
second derivatives of energy $E_{el}$ 
\begin{equation} 
\left.
\frac{\partial^2 E_{el}}
{
\partial {u}_{\alpha}^{\mu,0} 
\partial {u}_{\beta}^{\nu,{\bf h}}
}  
\right|_{\bf{u}=0},
\end{equation}
and therefore from Eq.~({\ref{E_elvar}})
is obvious that only the linear variation
of the electron density is needed.

\section{Variation of electron density}

The charge density $\rho({\bf r})$ is calculated from
\begin{equation}
\rho ({\bf r}) = \sum_{i} n_i \psi_{i}^*({\bf r}) \psi_{i}({\bf r}) ,
\end{equation}
where $\it{n_i}$ is the 
occupation number and $\psi_i$ are 
one-electron wavefunctions, the
variation of $\rho({\bf r})$
is expressed as
\begin{equation}
\delta \rho = 
\sum_{i} 
(n_i +\delta n_i)
(\psi_i^{0*} +\delta \psi_i^*)
(\psi_i^0 +\delta \psi_i)
- n_i \psi_i^{0*} \psi_i^{0}.
\end{equation} 
Having supposed that in the
non-metallic 
systems the 
$\delta n_i = 0$ 
one can write
\begin{equation}
\delta \rho = 
\sum_{i} 
n_i(\psi_i^{0*} \delta \psi_i 
+\delta \psi_i^*\psi_i^{0}
+\delta \psi_i^*\delta \psi_i).
\end{equation}
The variation of the wave 
function 
\begin{equation}
\delta \psi_i = \sum_j a_{ij} \psi_j^0 ,
\end{equation}
and the variation of the electron density 
can be written as
\begin{eqnarray} \label{varden}
\delta \rho &=&
\sum_{i} n_i
\sum_{j} (\psi_i^{0*} \psi_j^{0} a_{ij} 
+\psi_j^{0*} \psi_i^{0} a_{ij}^*)
\nonumber \\ &+&
\sum_{i} n_i
\sum_{jk} (\psi_j^{0*} \psi_k^{0} 
a_{ik} a_{ij}^*) .
\end{eqnarray}
The variation of the wave  function
is obtained from the Lippmann-Schwinger
 \cite{Ziman} equation
\begin{equation}
\mid{\psi_i}\rangle = \mid{\psi_i^0}\rangle
+ G_0 (\varepsilon_i) \delta V_{KS}~\mid{\psi_i}\rangle,
\end{equation}
where $G_0(\varepsilon)$ is the Green function
\begin{equation} \label{Green}
G_0(\varepsilon) =
\sum_i \frac{|\psi_i\rangle \langle \psi_i|}
{\varepsilon - \varepsilon_i},
\end{equation}
and $\delta V^{KS}$ is the variation of the total Kohn-Sham potential
\begin{equation}
\delta V^{KS} = \delta U + \delta V_c + \delta \mu_{xc}.
\end{equation}
With the use of Eqs.(\ref{varVc},\ref{var_mu})
it can be written as
\begin{equation}
\delta V^{KS} =
\delta U +
\int \frac{\delta \rho({\bf r'})}{|{\bf r-r'}|} d^3 r' +
\left (
\frac {\delta \mu_{xc}({\bf r})}{\delta \rho({\bf r})} 
\right )_0
\delta \rho({\bf r}).
\end{equation}
The elements $a_{ij}$ are then expressed as
\begin{equation} \label{elem}
a_{ij} =
\langle \psi_i^0 \mid \delta \psi_j \rangle 
=
\langle \psi_i^0 \mid G_0 (\varepsilon_i) \delta V_{KS} \mid \psi_j^0 \rangle.
\end{equation}
Having combined equations {(\ref{varden})} and {(\ref{elem})}
and applying operator $\partial_{\mu,\alpha}^{q}$
which contains the limit to the zero perturbation (Eq.(\ref{oper}))
we obtain integral equation
\begin{eqnarray} \label{selfint}
&&\partial_{\mu,\alpha}^{q}
\rho ({\bf r})
=
\sum_{i,j} n_i \psi_i^{0*}({\bf r})
\psi_j^0 ({\bf r})
\nonumber \\ &\times&
\langle \psi_i^0 \mid
G_0 (\varepsilon_i) 
\partial_{\mu,\alpha}^{q} 
(U + V_c + \mu_{xc}) 
\mid \psi_j^0 \rangle
\nonumber \\ &=&
\sum_{i,j} n_i \psi_i^{0*}({\bf r})
\psi_j^0 ({\bf r})
\langle \psi_i^0 \mid  
G_0 (\varepsilon_i) 
\partial_{\mu,\alpha}^{q} 
U 
\mid \psi_j^0 \rangle
\nonumber \\ &+&
\sum_{i,j} n_i \psi_i^{0*}({\bf r})
\psi_j^0 ({\bf r})
\langle \psi_i^0 \mid  
G_0 (\varepsilon_i)
\\ &\times&
{\left [
\int \frac{\partial_{\mu,\alpha}^{q} \rho ({\bf r}')}
{|r-r'|} d^3 r'
+
\left(
\frac{\delta \mu_{xc}({\bf r})}{\delta \rho({\bf r})}
\right )_0
\partial_{\mu,\alpha}^{q} \rho ({\bf r})
\right ]}
\mid \psi_j^0 \rangle,
\nonumber
\end{eqnarray}
which can be
rewritten into form
\begin{eqnarray} \label{self2}
\partial_{\mu,\alpha}^{q} \rho({\bf r})
&=&
\partial_{\mu,\alpha}^{q} \sigma({\bf r})
\\ &+&
\int \int \Lambda({\bf r},{\bf r'})
 \Gamma({\bf r'},{\bf r''})
d^3 r'
\partial_{\mu,\alpha}^{q} \rho({\bf r''})
d^3 r'',
\nonumber
\end{eqnarray}
where $\sigma$ and $\Lambda$ depend only
on eigenfunctions and eigenvalues of the
unperturbed state. The quantities from previous equation can be
expressed as
\begin{equation} \label{sigma}
\partial_{\mu,\alpha}^{q} \sigma({\bf r}) =
\sum_{i,j} \frac{n_i-n_j}{\varepsilon_i -\varepsilon_j}
\psi_i^{0*}({\bf r}) \psi_j^0 ({\bf r})
\langle \psi_j^0 \mid \partial_{\mu,\alpha}^{q}
U~\mid \psi_i^0 \rangle
\end{equation}
\begin{equation} \label{Lambda}
\Lambda({\bf r},{\bf r'}) =
\sum_{i,j} \frac{n_i-n_j}{\varepsilon_i-\varepsilon_j}
\psi_i^{0*}({\bf r}) \psi_j^0 ({\bf r})
\psi_j^{*0}({\bf r'}) \psi^0_{i}({\bf r'})
\end{equation}
\begin{equation} \label{Gamma}
\Gamma({\bf r},{\bf r'}) =
\frac{1}{|{\bf r''}-{\bf r'}|}
+\delta({\bf r'-r''})
\left (
\frac{\delta \epsilon_{xc}({\bf r''})}{\delta \rho({\bf r''})}
\right)_0.
\end{equation}
After expansion into plane waves the
integral equation (\ref{selfint}) yields
the set of algebraic equations. By solving
this set we obtain 
$\partial_{\mu,\alpha}^{q} \rho$ which are used
for calculation of the dynamical matrix.

The contribution of the variation of the
$E_{el}$ to the dynamical matrix
is given by the sum of contributions from
four terms on the right-hand side of the
Eq.~({\ref{E_elvar}}), i.e.
\begin{eqnarray}
D_{\alpha\beta}^{el}({\bf q},\mu\nu)
&=&
D_{\alpha\beta}^{el,1}({\bf q},\mu\nu)
+
D_{\alpha\beta}^{el,2}({\bf q},\mu\nu)
\nonumber \\ &+&
D_{\alpha\beta}^{el,3}({\bf q},\mu\nu)
+
D_{\alpha\beta}^{el,4}({\bf q},\mu\nu).
\end{eqnarray}
The contribution of the fifth term on the right-hand
side of the
Eq.~({\ref{E_elvar}}) is zero because
$\partial_{\mu,\alpha}^{q*}
\partial_{\nu,\beta}^{q}
\Theta (\delta U +\delta V_{c} +\delta\mu_{xc})
= 0$.

To calculate the term 
$D_{\alpha\beta}^{el,1}({\bf q},\mu\nu)$
is quite easy because it doesn't depend on
the variation of the charge density $\delta \rho$ 
but on the variation of the pseudopotential
$\delta U$ only. According to Eqs.~({\ref {E_elvar}})
one can write
\begin{equation} \label{D1}
D_{\alpha\beta}^{el,1}({\bf q},\mu\nu)
=
\sum_{v,{\bf k}}
\langle{v,{\bf k}}\mid 
\partial_{\mu,\alpha}^{q*}
\partial_{\nu,\beta}^{q}
U
\mid{v,{\bf k}}\rangle .
\end{equation}
More complicated is to calculate
$D_{\alpha\beta}^{el,2}({\bf q},\mu\nu)$.
According to the Eq.~({\ref{E_elvar}})
one can write 
\begin{eqnarray} \label{D2}
&& D_{\alpha\beta}^{el,2} ({\bf q},\mu\nu)
=
\sum_{v,c,k}
\frac{n_{v,k}-n_{c,k+q}}
{\varepsilon_{v,k}-\varepsilon_{c,k+q}}
\nonumber \\ &\times&
\langle{v,{\bf k}}\mid
\partial_{\mu,\alpha}^{q*}
(U +V_{c}+\mu_{xc})
\mid{c,{\bf k+q}}\rangle 
\nonumber \\ & \times &
\langle{c,{\bf k+q}}\mid 
\partial_{\nu,\beta}^{q}
(U + V_{c} +\mu_{xc})
\mid{v,{\bf k}}\rangle .
\end{eqnarray}
The sum over v and c in Eq.(\ref{D2})
runs over occupied and 
unoccupied bands respectively.
The $D_{\alpha\beta}^{el,3}({\bf q},\mu\nu)$
and $D_{\alpha\beta}^{el,4}({\bf q},\mu\nu)$
can be written in the forms
\begin{equation} \label{D3}
D_{\alpha\beta}^{el,3} (  {\bf q},\mu\nu)
=
-
\int 
\frac{
\partial_{\mu,\alpha}^{q*}
\rho({\bf r})
\partial_{\nu,\beta}^{q}
\rho({\bf r'})
}
{\bf |r-r'|}
d^3 r d^3 r' ,
\end{equation}
and
\begin{equation} \label{D4}
D_{\alpha\beta}^{el,4} ({\bf q},\mu\nu)
=
-
2
\int 
\left (
\frac{\delta \epsilon_{xc}({\bf r}) }{\delta \rho({\bf r})}
\right)_0
\partial_{\mu,\alpha}^{q*}
\rho({\bf r})
\partial_{\nu,\beta}^{q}
\rho({\bf r})
d^3 r .
\end{equation}

\section{Calculations}

The calculations were performed with all quantities expanded into plane
waves. The set of special ${\bf k}$-{\it points}
\cite{Baldereschi,Chadi-Cohen}
with $(\frac{1}{8},\frac{1}{8},\frac{1}{8})$ mesh was used for integration
over the Brillouin zone. It corresponds to 10 ${\bf k}$-{\it points}
for $\Gamma$ phonon and 128 ${\bf k}$-{\it points} for phonon with
general wave-vector.
The plane-wave set was selected for each ${\bf k}$ in the first Brillouin
zone such that among the reciprocal-lattice vectors ${\bf G}$ only those
are selected which have a kinetic energy
$E_{kin} = \frac{1}{2}({\bf k + G})^2$ less then $E_{cutoff} = 18 Ry$.
The sums over unoccupied states went over energies up to cutoff energy
$\varepsilon_{cutoff}= 9 Ry$, which corresponds to 350 unoccupied
bands approximately.

\section{Results}

The method presented in this paper was applied to silicon crystal
with lattice constant $a=0.543~ nm$. 
Both optical and acoustic branches of phonon frequencies in longitudinal and
transverse modes were obtained for $\Gamma, X$ and $L$ phonons.
Calculated frequencies together with values obtained by other authors and
experimental values are summarized in Table 1.
The values are expressed in $THz$.
\begin{table}
\caption{The calculated and experimental values of phonon
frequencies in the silicon crystal.}
\label{tab-frek}
{
\begin{tabular}[b]{lddd}
                              & $\Gamma$ &  $L$      &  $X$     \\
                              & $(0,0,0)$& $(\frac{1}{2},
                                          \frac{1}{2},
                                          \frac{1}{2})$
                                                     & $(1,0,0)$ \\ \hline
$ TO_{calc}\tablenotemark[1]$ &  15.9  &  15.8   &  15.1  \\
$ TO_{calc}\tablenotemark[2]
                        $ &  15.6  & $    - $  & $ -  $ \\
$ TO_{calc}\tablenotemark[3]
                        $ &  15.2  & $    - $  &  13.5  \\
$ TO_{calc}\tablenotemark[4]
                        $ &  15.7  & $    - $  & $  - $ \\
$ TO_{expt}\tablenotemark[5]
                        $ &  15.5  &  15.0     &  14.2  \\
\hline
$ LO_{calc}\tablenotemark[1]$ &  15.9  &  14.5     &  14.6  \\
$ LO_{calc}\tablenotemark[3]
                        $ &  $ -$  & $    - $  &  12.1  \\
$ LO_{expt}\tablenotemark[5]
                        $ &  15.5  &  12.8     &  12.5  \\
\hline
$ TA_{calc}\tablenotemark[1]$ &   0.0  &   6.2     &   7.3  \\
$ TA_{calc}\tablenotemark[6]
                        $ & $-$ &   3.0     &   4.1  \\
$ TA_{expt}\tablenotemark[5]
                        $ &   0.0  &   3.6     &   4.4  \\
\hline
$ LA_{calc}\tablenotemark[1]$ &   0.0  &  13.8     &  14.6  \\
$ LA_{expt}\tablenotemark[5]
                        $ &   0.0  &  11.7     &  12.5  \\ 
\end{tabular} 
}
{
\tablenotemark[1]{Phonons calculated in this work.} \\
\tablenotemark[2]{Reference[\onlinecite{Nielsen}].} \\
\tablenotemark[3]{Reference[\onlinecite{YIN}].}     \\
\tablenotemark[4]{Reference[\onlinecite{Hart}].}    \\
\tablenotemark[5]{Reference[\onlinecite{OPS}].}     \\
\tablenotemark[6]{Reference[\onlinecite{Wei}].}
}
\end{table}

\section{Discussion}
Having used the presented method a quite good agreement of the calculated
frequencies of optical phonons in silicon crystal and experimental
ones was reached.
We systematically increased the number of unoccupied states taken into
account and we have found that the increase over $\sim 300$ bands almost
doesn't change the results. Thus, the contribution from states with
energy $\varepsilon_i > \varepsilon_{cutoff}$ doesn't appear essential.
The biggest individual contributions to the dynamical matrix
come from unoccupied states with energy
near the Fermi energy. But the number of these states is small,
the density of states is approximately
$g(\varepsilon) \sim \varepsilon^{1/2}$, and so it can
be said that both dynamical matrix and phonon frequencies are integral
quantities where the influence of one state is not essential.
Any basis set for sums over unoccupied states is possible
but the eigenfunctions of the Kohn-Sham hamiltonian represent
the {\it natural choice} .

The presented method does not need time consuming calculations of
inverse matrix as in the {\it inverse dielectric matrix method}.
Method of Baroni {\it et al.} \cite{Baroni_et_al} is based on the
self-consistent scheme for perturbed eigenfunctions while our method
is based on the self-consistent equation (\ref{self2}) for the variation of
the electron density and only energies and eigenfunctions of
unperturbed states are needed.
King-Smith and Needs \cite{King-Smith_Needs}
used the Hellmann-Feynman forces on all atoms
in distorted crystal to construct the dynamical matrix.
The method of Gonze {\it et al.} \cite{Gonze} is based on variations of the
DFT total energy with respect to the first-order perturbations
of the wave functions.
In this paper, the second derivatives of energy approach is used.
The advantage of the presented method is that in contradiction
to the both, Baroni {\it et al.}
method and 
King-Smith and Needs
method, the 
iteration up to the self-consistency is needed only once for the unpertubed
state. In contradiction to the method of Gonze {\it et al.}, 
we don't need any additional
minimization of the total energy with respect to the first-order variations
of the wave function for each phonon, because all quantities
are calculated from quantities of unperturbed state.

Because of avoiding the iterations and minimization the presented
method is very suitable for massive parallelization of the computer
code.

\section{Conclusion}

The method for {\it ab initio} calculation of frequency and polarization
vector of phonons with an arbitrary wave-vector was presented.
The density functional theory within the pseudopotential
framework was used. As in other linear response methods, the presented
one uses the perturbation theory. In contradiction to the other
methods the presented one is based upon a direct calculation of the dynamical
matrix via second derivatives of the total crystal energy and no
calculation of the Hellman-Feynman forces is needed.

The presented method does not need time consuming calculations of
inverse matrix as in the {\it inverse dielectric matrix method}.
Another advantage is that in contradistinction
to the both Baroni {\it et al.} \cite{Baroni_et_al} method and
King-Smith and Needs \cite{King-Smith_Needs} method, no
iteration up to self-consistency for individual phonon is needed.
Similarly, we don't need any total energy minimization as in the
case of method of Rignanese, Michenaud and Gonze \cite{Rignanese}. 
The computing time saving due to the parallelization of the
computer code is significantly high for the presented method.

\end{document}